\documentclass[12pt]{article}
\usepackage{amsmath}
\usepackage{amssymb}
\usepackage{amsfonts}
\usepackage{ytableau}
\usepackage{cancel}
\usepackage{bm}
\usepackage{tikz}
\begin{document}
\def\be{\begin{equation}}
\def\bea{\begin{eqnarray}}
\def\ee{\end{equation}}
\def\eea{\end{eqnarray}}
\def\d{\partial}
\def\eps{\varepsilon}
\def\la{\lambda}
\def\b{\bar}
\def\nn{\nonumber \\}
\def\p{\partial}
\def\t{\tilde}
\def\h{{1\over 2}}
\def\cI{\mathcal{I}}
\def\hs{\hspace}
\makeatletter
\def\blfootnote{\xdef\@thefnmark{}\@footnotetext}  % for blank footnote
\makeatother

%\begin{document}

\title{\textbf{Notes on index of quantum integrability}}

\vspace{14mm}
\author{
	Jia Tian$^{1,3}$, Jue Hou$^{1}$,and Bin Chen$^{1,2,3}$\footnote{wukongjiaozi, houjue, bchen01@pku.edu.cn}
}
\date{}

\maketitle

\begin{center}
	{\it
		$^{1}$School of Physics and State Key Laboratory of Nuclear Physics and Technology,\\Peking University, No.5 Yiheyuan Rd, Beijing 100871, P.~R.~China\\
		\vspace{2mm}
		$^{2}$Collaborative Innovation Center of Quantum Matter, No.5 Yiheyuan Rd, Beijing 100871, P.~R.~China\\
		$^{3}$Center for High Energy Physics, Peking University, No.5 Yiheyuan Rd, Beijing 100871, P.~R.~China
	}
	\vspace{10mm}

\end{center}

\begin{abstract}
	A quantum integrability index was proposed in \cite{KMS}. It systematizes the Goldschmidt and Witten's operator counting argument \cite{GW} by using the conformal symmetry. In this work  we compute the quantum integrability indexes for the symmetric coset models ${SU(N)}/{SO(N)}$ and $SO(2N)/{SO(N)\times SO(N)}$. The indexes of these theories are all non-positive except for the case of ${SO(4)}/{SO(2)\times SO(2)}$. Moreover we extend the analysis to the theories with fermions and consider a concrete theory: the $\mathbb{CP}^N$ model coupled with a massless Dirac fermion. We find that the indexes for this class of models are non-positive as well.

\end{abstract}

\baselineskip 18pt

\newpage

\section{Introduction}
The study of integrability has a long history, which can date back to the time of the birth of Classical Mechanics\footnote{For a short history of integrability see \cite{YellowBook}.}. However the understanding of integrability is far from completion, particularly in the context of quantum field theories (QFT). The classical aspects of integrable QFT are usually described by the Lax operator formalism, which allows us to construct local or non-local classically conserved charges. The quantum aspects\footnote{For reviews of integrable structure in QFT see for example \cite{IntStructure}.} of integrable QFT are dictated by the S-matrix factorization and bootstrap \cite{SmatrixFactor}. Integrability itself is noble while proving integrability is always involved with sophisticated guesses and conjectures. The seminal works of \cite{Shankar:1977cm,Parke:1980ki} show that the factorization of S-matrix is a consequence of the existence of higher-spin quantum conserved currents. Nevertheless, the construction of  quantum conserved  currents is quite tricky, as the  classical conserved currents are often anomalous at the quantum level. 

In \cite{GW}, Goldschmidt and Witten (GW) proposed a \emph{sufficient} condition to prove the existence of quantum conserved currents. By enumerating all the possible local operators which can appear in the anomaly of the classical conservation laws one can tell whether there exist quantum conserved currents. Even though 
the GW argument is clear,  the complexity in counting the possible local operators in practice by the brutal-force method goes wild quickly. Recently, Komatsu, Mahajan and Shao (KMS) \cite{KMS} systematized the counting, and introduced a quantum integrability index $\cI(J)$ for each spin $J$, which we call the KMS index, to characterize the existence of the quantum higher-spin conserved currents. It is a lower bound on the number of quantum conserved currents of spin $J$. If the KMS index $\cI(J)$ is positive, it implies the existence of the quantum conserved currents of the spin $J$. One remarkable feature of the KMS index is that it is usually defined at the UV fixed point of the sigma-model, but it is invariant under conformal perturbation around a conformal field theory fixed point. This allows us to use the conformal symmetry to enumerate the gauge invariant operators according to their scaling dimensions in a systematical way such that the computation of the index is feasible. In \cite{KMS}, the indexes of the higher spin currents for the $\mathbb{CP}^{N-1}$ model, the $O(N)$ model and the flat sigma model $\frac{U(N)}{U(1)^n}$ were computed. 

In this note, we would like to compute the KMS index for some other quantum integrable coset models, including the $SU(N)/SO(N)$ model,  the $SO(2N)/SO(N)\times SO(N)$ model and the $\mathbb{CP}^N$ model coupled with a Dirac fermion. We find that the KMS indexes of higher spins in these models are all non-positive except for $\cI(4)$ in the  ${SO(4)}/{SO(2)\times SO(2)}$  model. 

The organization of the paper is as follows. In section 2, we review the GW argument and the KMS quantum integrable index. For a clear illustration we focus on a concrete example, $O(N)$ model. In section 3, we compute the KMS index for the coset models $SU(N)/SO(N)$ and $SO(2N)/SO(N)\times SO(N)$. These two models are conjectured to be quantum integrable. Also in section 3, we consider the models with fermions and show how to generalize the KMS index. We summarize our results in section 4.

\section{GW argument and KMS  index}
\renewcommand{\theequation}{2.\arabic{equation}}
\setcounter{equation}{0}
In this section we briefly review the Goldschmidt and Witten's arguments for quantum integrability \cite{GW} and the quantum integrability index, introduced by Komatsu, Mahajan and Shao \cite{KMS}. We will take the $O(N)$ model to  elaborate the analysis. % which systematically implements the GW's argument instead of a brute force treatment.

\bigskip

\noindent{\textbf {GW argument}}

 In \cite{GW}, Goldschmidt and Witten proposed a sufficient condition to diagnose the conservation of quantum higher-spin currents in two dimensional sigma models. Their criterion is based on an operator counting analysis in sigma models. Consider a two dimensional sigma model with classical conserved current satisfying 
\bea 
\p_-\mathcal{J}_+^{cl}=0.
\eea 
Quantum mechanically, the classical symmetry may be broken such that  the conservation equation is modified to
\bea \label{GWB}
\p_-\mathcal{J}_+^{qu}=A,
\eea 
where the anomalous term $A$ is a  local operator with proper conformal dimension. However, if $A$ can be written as a total derivative as
\bea 
A=\p_+ B_-+\p_- B_+,
\eea 
then one may redefine the current as
\bea 
(\mathcal{J}_+^{qu},\mathcal{J}_-^{qu}):= (\mathcal{J}_+^{cl}-B_+,\mathcal{J}_-^{cl}-B_-)
\eea 
such that the redefined current is conserved quantum mechanically. The GW criterion is that if the number of $A$-type operators is less than the number of $B$-type operators then the quantum higher-spin current is conserved. 
As an example \cite{GW} we consider the $O(N)$ $\sigma$ model whose action is given by
\bea \label{Naction}
\mathcal{L}=\frac{1}{2\alpha}\p_\mu \vec{n}\cdot\p_\mu \vec{n},\quad |\vec{n}|=1.
\eea 
The theory is classically conformal invariant, and it has conserved currents of even spin building from the stress tensor. The stress tensor of the theory is $T_{++}=\p_{+}\vec{n}\cdot \p_{+}\vec{n}$. Due to the fact that $\p_- T_{++}=0$, the currents $J_n =(T_{++})^n$ is conserved classically. Let us consider the classical conserved spin-4 current $T_{++}^2$ and then \eqref{GWB} reads
\bea 
\p_-[(\vec{n}_+\cdot \vec{n}_+)^2]=A,\quad \mbox{where}\quad \vec{n}_\pm:=\p_{\pm}\vec{n}.
\eea 
To construct the $A$-type and $B$-type local operators, we first find the building blocks, a list of fundamental independent local operators called the letters.  The requirement that the operators should be $O(N)$ invariant implies that the vector index of one $\vec{n}$ must contract with the one of  another $\vec{n}$ to get a $O(N)$ singlet. Due to the constraint $|\vec{n}|=1$, we can claim $\vec{n}$ and $\p\vec{n}\cdot \vec{n}$ are not in the list. On the other hand, the equation of motion (EOM) of the model is
\bea 
\p_+\p_- \vec{n}=-\vec{n} \p_+ \vec{n}\cdot \p_- \vec{n} ,
\eea 
which implies that the letters can not have cross derivatives.
Therefore the possible letters are
\bea \label{ONletter}
P^{pq}_{++}=\p_+^p \vec{n}\cdot \p_+^q \vec{n},\quad P^{pq}_{+-}=\p_+^p \vec{n}\cdot \p_-^q \vec{n},\quad P^{pq}_{--}=\p_-^p \vec{n}\cdot \p_-^q \vec{n},
\eea
with conformal dimensions
\bea 
h_{++}^{pq}=(p+q,0),\quad h_{+-}^{pq}=(p,q),\quad h_{--}^{pq}=(0,p+q), \quad p,q\geq 1.
\eea 
Since the conformal dimension of $A$ is $h_A=(h,\bar{h})=(4,1)$
the only possible A-type operators are
\bea 
A_1=P_{+-}^{4,1},\quad A_2=P_{+-}^{1,1}P_{++}^{2,1},\quad A_3=P_{+-}^{2,1}P_{++}^{1,1}.
\eea 
The conformal dimension of $B_+$ and $B_-$  are $h_+=(4,0)$ and $h_-=(3,1)$, respectively. So they can be
\bea 
&&B_{+1}=P_{++}^{1,3},\quad B_{+2}=P_{++}^{2,2},\quad B_{+3}=P_{++}^{1,1}P_{++}^{1,1},\nn
&&B_{-1}=P_{+-}^{3,1},\quad B_{-2}=P_{+-}^{1,1}P_{++}^{1,1}.
\eea 
It seems that there are five $B$-type operators, but that is not true because we have not imposed the EOM. In other words, these  $B$-type operators are not independent, considering the EOM. To remove the redundancy we have to rewrite $\p_\pm B_{\pm}$ in terms of A:
\bea 
&&\p_- B_{+3}=0,\quad \p_-B_{+1}=-4 A_2-2A_3,\quad \p_-B_{+2}=-2A_2+2A_3,\\
&&\p_+B_{-1}=A_1+3A_2,\quad \p_+B_{-2}=A_3+2A_2.
\eea 
Therefore, there are only three independent $B$-type operators remaining after imposing the EOM. It implies that $A$ can always be written as a total derivative so that the spin-4 current is conserved even at the quantum level. 

\bigskip

\noindent{\textbf {KMS index}}

Following the GW argument, the authors in \cite{KMS} proposed the index
\bea \label{KMSIndex}
\mathcal{I}(j)= \#(\mathcal{J}^{cl}_j)-[\#(A)-\#(B)].
\eea 
where $\mathcal{J}^{cl}_j$ are classically conserved currents of spin $j$. 
If $\mathcal{I}(j)>0$, then it is guaranteed that there exit at least $\mathcal{I}(j)$ quantum conserved currents of spin $j$. 

In the brutal-force counting method, we have shown the most cumbersome step is to remove the redundancy  in the counting of $B$-type operators, due to the on-shell equation of motion. Noticing  that the difference $A-B$ defines the set
\bea
\mathcal{C}=\{A\}-\{B\}=\frac{\{A\}}{\mbox{EOM} \times \mbox{IBP}}. 
\eea
Here IBP stands for the total derivative terms as known as  Integration By Part. 
The set $\mathcal{C}$ can be interpreted as the set of local operators with proper quantum numbers after considering the EOM and IBP. This kind of object has a clear analogue in effective field theory (EFT) known as the operator bases \cite{OperatorB}. The crucial idea here is that as the index is invariant under conformal deformation, we can study the index at the UV fixed point where we can organize all the local operators with respect to the conformal multiplets schematically denoted as
\bea 
\{\mathcal{O},\p \mathcal{O},\p^2\mathcal{O},\dots \}
\eea 
As a result, the partition function $Z$ for all the independent local operators (the letters)  has an expansion with respect to the conformal group characters $\tilde{\chi}_{\Delta,j}$ labeled by the conformal dimension $\Delta$ and the spin $j$:
\bea 
Z(q,x)\equiv\sum_{\mathcal{O}}q^{\Delta_\mathcal{O}}x^{j_\mathcal{O}}=\sum_{\Delta ,j}c(\Delta, j)\tilde{\chi}_{\Delta ,j}.
\eea 
Applying the orthogonal property of the character, the KMS index \eqref{KMSIndex} for the spin\footnote{The spin $j$ has to be an integer in order to have an inversion formula.} $j$ could be computed by using an inversion formula \cite{KMS}
\bea \label{KMSIndexInt}
\mathcal{I}(j)=c(j,j)-c(j+1,j-1)=-\int  Z(q,x)~\chi_{j+1,j-1}^\star d\mu_{q,x},
\eea
where a dual character $\chi^\star_{\Delta,j}(q,x)$ is defined as
\bea 
\chi^\star_{\Delta,j}(q,x)=\tilde{\chi}_{\Delta,j}(1/q,1/x).
\eea 

Let us revisit the $O(N)$ model with this approach. The single-letter characters corresponding to the letters \eqref{ONletter} is
\bea 
\chi(q,x)&=&=\sum_{m\geq 1,n\geq m}(q^{m+n}x^{m+n}+q^{m+n}x^{-m-n})+\sum_{m,n\geq 1}q^{m+n}x^{m-n}\nn
&=&\frac{qx}{1-qx}\frac{qx^{-1}}{1-qx^{-1}}+\frac{1}{1-qx}\frac{q^2 x^2}{1-q^2 x^2}+\frac{1}{1-q/x}\frac{q^2/ x^2}{1-q^2/ x^2}
\eea 
The multi-letter partition function is given by the plethystic exponential \cite{OperatorB}:
\bea 
Z(q,x)=\mbox{PE}(\chi)=\exp(\sum_{m=1}\frac{1}{m}\chi(q^m,x^m)).
\eea 
To compute the quantum index $\mathcal{I}(4)$ we need  the character
\bea 
\chi_{5,3}=q^5x^3\sum_{n,m}q^{n+m}x^{n-m}=\frac{q^5 x^3}{(1-qx)(1-qx^{-1})},\quad \chi_{5,3}^\star(q,x)=\chi_{5,3}(1/q,1/x).
\eea 
and the measure in the space $(q,x)$ 
\bea 
\int \mu_{q,x}=\oint \frac{dq}{2\pi i q}\oint \frac{dx}{2\pi i x}(1-qx)(1-q^{-1}x)(1-q^{-1}x^{-1})(1-qx^{-1}).
\eea 
Substituting into \eqref{KMSIndex}, one can find $\mathcal{I}(4)=1$ which matches the results from  brutal force method. We conclude this section by listing other KMS indexes for the $O(N)$ model:
\bea \label{ONindex}
&&\mathcal{I}(4)=1,\hs{2ex}\mathcal{I}(6)=1,\hs{2ex}\mathcal{I}(8)=0,\nn
&&\mathcal{I}(10)=-5,\hs{2ex}\mathcal{I}(12)=-15,\hs{2ex}\mathcal{I}(14)=-43\dots
\eea 
Thus, there also exists a spin-6 quantum conserved current, as predicted in \cite{KMS}.

%The KMS index can be computed for other coset sigma models. Some examples had been studied in \cite{KMS}. In this section we study some other

\section{Coset models}
\renewcommand{\theequation}{3.\arabic{equation}}
\setcounter{equation}{0}

 The sigma models on homogeneous spaces also known as symmetric coset models are important examples of classical integrable field theory \footnote{A recent review can be found in \cite{Zarembo:2017muf}. For an integrable but not symmetric coset model see \cite{Flag}.}. Applying the operator counting techniques developed for EFT \cite{OperatorB}, KMS proposed a systematic way to compute the integrability index for the coset sigma models, which are  \emph{not} necessary to be symmetric. 

\bigskip

\noindent{\textbf {KMS index for cosets}}
Consider a coset $G/H$ with the associated Lie algebra orthogonal decomposition
\bea 
\mathfrak{g}=\mathfrak{h}\oplus\mathfrak{k},
\eea 
where $\mathfrak{h}$ and $\mathfrak{k}$ represent the elements in subalgebra and coset, respectively. Introducing the left-invariant one-form 
\bea 
&&j_\mu(x)\equiv g^{-1}(x)\p_\mu g(x) ,\quad g(x)\in G,\quad j_\mu(x)\in\mathfrak{g},\eea 
and its decomposition
\bea
j_\mu(x)=a_\mu(x)+k_\mu(x), \quad a_\mu(x)\in \mathfrak{h},\quad k_\mu(x)\in \mathfrak{k},
\eea
 the action of the sigma model can be written as
 \bea \label{CosetAction}
 S=\frac{R^2}{2}\int \mbox{Tr}[k_\mu(x)k^\mu(x)].
 \eea 
The coset model has the local symmetry:
\bea 
g(x) \rightarrow g(x)h(x)^{-1},\quad h(x)\in H
\eea 
and a global symmetry:
\bea 
g(x)\rightarrow g' g(x),\quad g'\in G.
\eea 
The local operators can be built from $g,k_\mu(x)$ and their covariant derivatives $D_\mu$ which is defined as
\bea 
D_\mu: \p_\mu+a_\mu.
\eea 
By imposing EOM and the flatness condition of the left-invariant one-form we can find the complete set of global $G$-symmetry invariant letters 
\bea \label{KMSLetter}
k_+^{(n)}\equiv (D_+)^n k_+,\quad k_-^{(n)}\equiv (D_-)^{n}k_-,
\eea 
where the light-cone coordinates have been used. All the letters under the $H$ gauge transformation transform as $h k h^{-1}$.
From $g$ and $k_\mu(x)$ we can built the Noether currents of the global $G$ symmetry
\bea
J_\mu(x):=g(x)k_{\mu}(x)g^{-1}(x).
\eea 
Using the Noether currents  we can find a set of $H$-symmetry invariant letters 
\bea \label{GroupLetter}
J_+^{(n)}\equiv (\tilde{D}_+)^n J_+,\quad J_-^{(n)}\equiv (\tilde{D}_-)^{n}J_-,\quad \tilde{D}_\mu\equiv\p_\mu+[J_\mu,\cdot]. 
\eea  
Since $\mbox{Tr}(\tilde{D}J^m)=\mbox{Tr}(Dk^m)$, one may think that all the $H$-invariant local operators on $\mathfrak{k}$ can be constructed from the $G$-invariant local operators on $\mathfrak{g}$. This is not true because the representation $r$ of $\mathfrak{h}$ which the vector space $k\in \mathfrak{k}$ forms is reducible and we can decompose $r$ into the irreducible representations of $H$: $r=\oplus_i r_i$. From each $r_i$ we can construct a set of  gauge invariant local operators.  Therefore the set of letters \eqref{GroupLetter} is not complete. 

In order to construct gauge invariant operators, KMS introduced auxiliary parameters which they call fugacities for the representations, and performed the Haar integration over the group $H$. As a result, the single-letter character is given by
\bea 
\chi(q,x,y_i)\equiv\sum_{n=0}^{\infty}q^{n+1}(x^{n+1}+x^{-n-1})\chi_R(y_i)=(\frac{xq}{1-xq}+\frac{x^{-1}q}{1-x^{-1}q})\chi_R(y_i),
\eea 
and the multi-letter partition function is similarly given by the plethystic exponential
\bea 
Z(q,x)=\int d\mu_{H} Z(q,x,y_i),\quad \mbox{with} \quad Z(q,x,y_i)=\mbox{PE}(\chi(q,x,y_i)).
\eea 
From the KMS index point of view, the quantum integrability is totally determined by the representation $R$ and the measure $d\mu_H$. 
 When the representation $R$ is trivial i.e. $\chi_R=1$, the KMS index vanishes. It is not hard to verify this fact numerically. For example, the index for the spin-4 current is given by
\bea 
\mathcal{I}(4)&=&-\frac{1}{24} (\chi (1)-1) [\chi (1)^4+8 \chi (1)^3+6 (\chi (2)+2) \chi (1)^2\nn
&~&+8 \chi (3) \chi (1)+3 \chi (2) (\chi (2)+4)-8 \chi (3)+6 \chi (4)],\quad \chi(m)\equiv \chi(y_i^m).
\eea 
 It is obviously vanishing for the trivial representation. We can also understand it in an intuitive way. For any high-spin conserved current $J_+^n$
there exist a $A$-type operator $A^{(n)}=k_-^{(1)}J_+^n$. Because no cross derivatives can appear there is no $B$-type operators then the KMS indexes have to vanish. But we want to stress that the vanishing of KMS indexes does not mean the theory is not integrable. Instead we should think that in this situation GW argument fails and in order to examine the quantum integrability we need some other tools or criteria

Let us revisit the $O(N)$ model which can be viewed as the coset model $\frac{SO(N)}{SO(N-1)}$. The currents $k_\mu$ form a vector representation of $SO(N-1)$. For simplicity, we assume $N-1$ to be even then the character of the vector representation is given by
\bea \label{ONindex2}
\chi_R=\sum_{i=1}^{(N-1)/2}(y_i+y_i^{-1}),
\eea 
and the Haar measure is given by
\bea 
d\mu(y)=\prod_{i}\frac{dy_i}{2\pi i y_i}\prod _{i<j}(1-y_iy_j)(1-\frac{y_i}{y_j}).
\eea 
Using the formula \eqref{KMSIndexInt}, we find the following results:
\bea\label{ON}
\begin{array}{|c|c|c|c|c|c|c|}
\hline
~&\mathcal{I}(4)&\mathcal{I}(6)&\mathcal{I}(8)&\mathcal{I}(10)&\mathcal{I}(12)&\mathcal{I}(14)\\
\hline
N=3&0&-1&-5&-15&-33&-75\\
N=5&1&0&-2&-9&-27&-71\\
N=7&1&1&0&-5&-15&-43\\
N=9&1&1&0&-5&-15&-43\\
\hline
\end{array}
\eea
The observation is that when $N$ is small the integrability indexes depend on $N$  but they become stable when $N\geq 7$ and the stabilized values coincide with results \eqref{ONindex}. Our calculations \eqref{ONindex} and \eqref{ONindex2} show that the two descriptions \eqref{Naction} and \eqref{CosetAction} of the $O(N)$ model are only equivalent for large enough $N$. The discrepancy between two kinds of counting for small $N$ is subtle. We believe that the  counting in the coset description is reliable. The subtlety is that in the description \eqref{Naction} after imposing the constraints
\bea 
\vec{n}\cdot \p_+ \vec{n}=\vec{n}\cdot \p_- \vec{n}=0,
\eea 
the vectors $\p_+ \vec{n}$ and $\p_- \vec{n}$ are orthogonal to $\vec{n}$ such that the two $N$-vectors live in a  $(N-1)$ dimensional subspace. It implies that the constraint $|\vec{n}|=1$ has not been fully imposed in the counting. To impose the constraint completely we should introduce the projected coordinates $\vec{\xi}=(\xi^1,\dots,\xi^{N-1})$ defined by
\bea 
n^i=\frac{2\xi^i}{1+|\xi|^2},\quad i=1,\dots,N-1,\quad n^N=\frac{1-|\xi|^2}{1+|\xi|^2}.
\eea 
The new letters $\p_\pm^{(n)}\xi^i$ are then in one-to-one map with  $k_\pm^{(n),i}$.

The discrete symmetry plays an important role for the quantum integrability. For example, for the parity-symmetric theories, the existence of only one local higher-spin conserved current will guarantee the quantum integrability. For the models with discrete symmetry, the KMS indexes must be improved by imposing the discrete symmetry. In this case,   we can modify the partition function by gauging the discrete symmetry group $\tilde{G}$ as \cite{KMS}
\bea \label{Discrete}
\tilde{Z}(q,x)\equiv=\frac{1}{|\tilde{G}|}\sum_{i}Z_{\tilde{g}_i},\quad \tilde{g}_i\in\tilde{G},\quad Z_{\tilde{g}_i}=\sum_{\mathcal{O}}[\tilde{g}_iq^{\Delta_\mathcal{O}}x^{j_\mathcal{O}}].
\eea 
Imposing the discrete $Z_2$ charge-conjugation symmetry, the KMS indexes of the $O(N)$ model become\footnote{In \cite{KMS}, the indexes $\mathcal{I}(4)$, $\mathcal{I}(6)$ and $\mathcal{I}(8)$ have been computed.}
\bea\label{ONZ2}
\begin{array}{|c|c|c|c|c|c|c|}
\hline
~&\mathcal{I}(4)&\mathcal{I}(6)&\mathcal{I}(8)&\mathcal{I}(10)&\mathcal{I}(12)&\mathcal{I}(14)\\
\hline
N&1&1&0&-4&-11&-30\\
\hline
\end{array}
\eea
independent of $N$. Comparing with the results without imposing the discrete symmetry, we see that the indexes of spin $4$ and $6$ are always positive, and the indexes of  higher spin are larger than the one without discrete symmetry. %Comparing with previous results \eqref{ONindex}, the indexes only match for small $J$. 

\bigskip

In the next section, we will use this strategy to study a few classical integrable models. For coset models, the crucial step is to identify the representation of $k_\mu$ with respect to the subgroup. That is involved with a representation decomposition problem. Since we only need the character of the representation we solve the problem in the following way. Firstly we separate the normalized generators $\{T_M\}$ of the group into the subgroup part $\{T_a\}$ and the coset part $\{T_\alpha\}$. Then we parameterize the subgroup element as
\bea 
h=\exp(ix^a T_a)
\eea 
so that the representation $R$ is given by
\bea \label{Rep}
R_{\alpha\beta}=\mbox[T_\alpha h T_{\beta}h^{-1}].
\eea 
In the end we express the character of $R$ in terms of the eigenvalues of $h$ which are our auxiliary parameters of fugacities.

\section{Applications}
\renewcommand{\theequation}{4.\arabic{equation}}
\setcounter{equation}{0}
\subsection{Cosets $SU(N)/SO(N)$}
The exact S-matrices for the sigma models on the spaces $SU(N)/SO(N)$ and $SO(2N)/SO(N)\times SO(N)$ were derived  in \cite{Fendley:2000bw} where the author also showed when the $\theta$ term equals $\pi$ the sigma models have stable low-energy fixed points corresponding to $SU(N)_1$ and $SO(2N)_1$ Wess-Zumino-Witten (WZW) models. The quantum integrability of these two models relies on the fact that \emph{non-local} charges survive quantization \cite{Evans:2004ur}. In this and next sections, we examine the conservation of \emph{local} higher-spin currents using the KMS index. 

To identify the generators  of the subgroups $SO(N)$ for the symmetric cosets $SU(N)/SO(N)$ we can solve the following equations \cite{SUSO}
\bea 
&&T_a\Sigma_0+\Sigma_0T_a^T=0,\quad T_\alpha\Sigma_0-\Sigma_0T_\alpha^T=0,
\eea 
where $\Sigma_0$ ia an $N\times N$ complex symmetric matrix that satisfies $\Sigma_0^\dagger \Sigma=|c|^2 I$ for some complex number $c$.  Using the Gell-Mann matrices as the generators of $SU(3)$, one can find that 
\bea 
&&T_a:\quad \{\frac{1}{2}(\lambda_1-\lambda_6), \frac{1}{2}(\lambda_2-\lambda_7),\frac{1}{2}(\lambda_3+\sqrt{3}\lambda_8)\},\nn
&&T_\alpha:\quad \{(\lambda_1+\lambda_6)/2,(\lambda_2+\lambda_7)/2,\frac{1}{2}\sqrt{\frac{3}{2}}(\lambda_3-\lambda_8/\sqrt{3}),\lambda_4/\sqrt{2},\lambda_5/\sqrt{2}\},
\eea 
where we have normalized the generators as $\mbox{Tr}[T_\alpha T_\beta]=\delta_{\alpha\beta}$. The character of the representation \eqref{Rep} is 
\bea 
\chi_R(N=3)=1+y+y^{-1}+y^2+y^{-2}.
\eea 
Taking a higher dimensional analog of the defining generators $\lambda_i,~i=1,\dots,15$. we find the decomposition of the normalized generators
\bea 
&&T_a:\quad \{\lambda_1-\lambda_{13},\lambda_2-\lambda_{14},\lambda_4-\lambda_{11},\lambda_5-\lambda_{12},\frac{\lambda_3}{\sqrt{2}}+\frac{\lambda_8}{\sqrt{6}}+2\frac{\lambda_{15}}{\sqrt{3}},\frac{\lambda_3-\sqrt{3}\lambda_8}{\sqrt{2}}\}/2,\nn
&&X_b:\quad {\{\lambda_1+\lambda_{13},\lambda_2+\lambda_{14},\lambda_4+\lambda_{11},\lambda_5+\lambda_{12},\sqrt{2}\lambda_{6,7,9},\lambda_3+\frac{\lambda_8}{\sqrt{3}}-\sqrt{\frac{2}{3}}\lambda_{15}}\}/2. \nn
\eea 
The corresponding character of the representation \eqref{Rep} is given by
\bea 
\chi_R(N=4)&=&(1+y_1y_2+1/(y_1y_2))(1+y_1/y_2+y_2/y_1)\nn&=&1+y_1^2+y_2^2+y_1^{-2}+y_2^{-2}+y_1y_2+y_1^{-1}y_2^{-1}+y_1^{-1}y_2+y_1y_2^{-1}.
\eea 
The  observation is that the representation $R$ is  the totally symmetric representation $[2,0\dots,0]$. Using the expressions the Haar measures for the groups $SO(N)$ \cite{OperatorB}, we get 
\bea 
&&N=3:\quad \mathcal{I}(4)=-3,\quad \mathcal{I}(6)=-7,\quad \mathcal{I}(8)=-34\\
&&N=4:\quad \mathcal{I}(4)=-3,\quad \mathcal{I}(6)=-19,\quad \mathcal{I}(8)=-100.
\eea 
The negative indexes imply that the GW argument fails. 

We now proceed to take care of the discrete symmetry. Imposing the charge conjugation discrete symmetry extends the gauge group from $SO(N)$ to $O(N)$. The orthogonal group $O(N)$  consists of two connected components: $O_+(N)=SO(N)$ and the parity-odd component $O_-(N)$. A general element $g_-\in O_-(N)$ is connected to an element $g_+\in SO(N)$ through a parity transformation $\sigma$ in the form $g_-=g_+\sigma$. For odd $N$, the parity transformation can be chosen to commute with the rotations due to $O(2r+1)=SO(2r+1)\times Z_2$ so that $\sigma_{[1]}=-I$. Noticing  $\sigma_{[2]}=(-1)^2I$ and $d\mu_-=d\mu_+$ we conclude that the $Z_2$ symmetry does not change the KMS index for odd $N$ cases. For even $N$ case, because of $O(2r)=SO(2r)\rtimes Z_2$, the parity transformation $\sigma$ does not commute with the rotation anymore. The results\footnote{For example, see the appendix of \cite{OperatorB}.} of the representation theory is that the general irreducible representation of $O(2r)$ are labeled by $l=(l_1,\dots,l_r)$ with $l_1\geq \dots l_r \geq 0$,
\bea 
&& l_r>0:\quad R^{O(2r)}_{l_1,\dots,l_{r-1},l_r}=R^{SO(2r)}_{l_1,\dots,l_{r-1},l_r}\oplus R^{SO(2r)}_{l_1,\dots,l_{r-1},-l_r}\nn
&&l_r=0:=\quad R^{O(2r)}_{l_1,\dots,l_{r-1},0}=R^{SO(2r)}_{l_1,\dots,l_{r-1},l_r},
\eea 
with the corresponding characters
\bea 
&& l_r >0:\quad \chi_l^+(x)=\chi_{(l_1,\dots,l_r)}(x)+\chi_{(l_1,\dots,-l_r)}(x),\quad \chi_l^{-}(\tilde{x})=0,\nn
&& l_r=0:\quad \chi_l^+(x)=\chi_l(x),\quad \chi_l^{-}(\tilde{x})= \chi^{Sp(2r-2)}_{l_1,\dots,l_{r-1}}(\tilde{x}).
\eea 
At the same time taking the measure  $d\mu_-=d\mu_{Sp}$ one can find the KMS index with \eqref{KMSIndexInt}. In the example of $N=4$, we obtain
\bea 
N=4: \quad \mathcal{I}_-(4)=-1,\quad \mathcal{I}_-(6)_-=-7,\quad \mathcal{I}_-(8)=-26.
\eea 
In the end combining the two components with \eqref{Discrete} gives total KMS indexes
\bea 
N=4: \quad \mathcal{I}_t(4)=-2,\quad \mathcal{I}_t(6)=-13,\quad \mathcal{I}_t(8)=-63.
\eea 
So the KMS index does not predict the existence of the quantum conserved spin-4 currents or any other higher-spin currents for these coset models. This is actually true for other even $N$. 
In short,  the high-spin KMS indexes for the cosets $SU(N)/SO(N)$ are all negative for all $N$, no matter $N$ is odd or even.

\subsection{Cosets $SO(2N)/SO(N)\times SO(N)$}
The symmetric cosets $SO(2N)/SO(N)\times SO(N)$ are known as the Grassmannians. We  present the details for the low-rank examples, and then conclude for general $N$. 

Let us start with the lowest rank case
\bea \label{Grass1}
SO(4)/SO(2)_1\times SO(2)_2.
\eea  
We will use the defining normalized generators for the orthogonal groups. In this case, the subgroup corresponds to the Cartan subgroup spanned by $(T_{12},T_{34})$. The character of the representation $R_{ab}$ are
\bea 
\chi_R=(y_1+y_1^{-1})(y_2+y_2^{-1}),
\eea 
and the corresponding measure is 
\bea 
d\mu=\frac{dy_1}{2\pi i y_1}\frac{dy_2}{2\pi i y_2}.
\eea 
The product form of the character is due to the fact the coset is in the bi-fundamental representation: $R=[1]_1\otimes [1]_2$.
A direct calculation gives the KMS indexes
\bea 
&&\mathcal{I}(2)=2,~\mathcal{I}(4)=-7,~\mathcal{I}(6)=-30,~\mathcal{I}(8)=-116,\dots.
\eea 
However the Grassmannian \eqref{Grass1} is basically two copies of $\mathbb{CP}^1$, so we expect that the KMS indexes can be improved by imposing discrete symmetries. Because locally $SO(4)\sim SU(2)_1\times SU(2)_2$, the parity group is
\bea 
Z_2\times Z_2: \quad \{I,\sigma\otimes I,I\otimes \sigma,\sigma\otimes \sigma\},\quad \sigma=\begin{bmatrix}
	1&0\\
	0&-1
\end{bmatrix} .
\eea
Apart from this there is another $Z_2^\tau$ symmetry which swaps the two $SU(2)$'s whose generator is
\bea 
\tau=\left(
\begin{array}{cccc}
 1 & 0 & 0 & 0 \\
 0 & 0 & 1 & 0 \\
 0 & 1 & 0 & 0 \\
 0 & 0 & 0 & 1 \\
\end{array}
\right),\quad \tau\sigma_1=\sigma_2\tau.
\eea 
Multiplying  the elements in $Z_2\times Z_2$ by $\tau$, we can generate more elements:
\bea 
&&\tau\sigma_1=\left(
\begin{array}{cccc}
 1 & 0 & 0 & 0 \\
 0 & 0 & -1 & 0 \\
 0 & 1 & 0 & 0 \\
 0 & 0 & 0 & -1 \\
\end{array}
\right),\hs{2ex} \tau\sigma_2=\left(
\begin{array}{cccc}
 1 & 0 & 0 & 0 \\
 0 & 0 & 1 & 0 \\
 0 & -1 & 0 & 0 \\
 0 & 0 & 0 & -1 \\
\end{array}
\right), \hs{2ex}\tau\sigma_{12}=\left(
\begin{array}{cccc}
 1 & 0 & 0 & 0 \\
 0 & 0 & -1 & 0 \\
 0 & -1 & 0 & 0 \\
 0 & 0 & 0 & 1 \\
\end{array}
\right).\nn
\eea 
Averaging over the full discrete group $Z_2\times Z_2\times Z_2^\tau$ we end up with the final KMS indexes 
\bea 
\mathcal{I}(2)=1,~\mathcal{I}(4)=1,~\mathcal{I}(6)=-1,~\mathcal{I}(8)=-10,\dots
\eea 
Indeed the spin-4 quantum conserved charge is recovered. 

\bigskip

For the higher rank case, the letters $k^{i\alpha}_\mu$ still transform in the bi-fundamental representation of $SO(N)_1\times SO(N)_2$ therefore the character is also given by a product of two individual characters:
\bea 
\chi_R=\chi_1(y_i)\chi_2(y_\alpha),\quad R=r_{[1]_1}\otimes r_{[1]_2}.
\eea 
We find that all the higher-spin KMS indexes are negative. Imposing the parity group $Z_2\times Z_2$ will not help. For examples, one can obtain
\bea 
&&N=3:\quad \mathcal{I}(2)=1,\quad \mathcal{I}(4)=0,\quad \mathcal{I}(6)=-6,\quad \mathcal{I}(8)=-43,\\
&&N=4:\quad \mathcal{I}(2)=1,\quad \mathcal{I}(4)=0,\quad \mathcal{I}(6)=-6,\quad \mathcal{I}(8)=-45.
\eea 
When $N>4$, the subgroups are not Abelian and the representation $R$ is not reducible so that we do not have the $Z^\tau_2$ symmetry anymore. Therefore, we conclude that KMS index fails to predict the existences of the quantum conserved higher-spin currents\footnote{Here we have not considered the Pfaffian currents which could give a spin-$N$ conserved currents \cite{Evans:2004ur}. } for the coset models $SO(2N)/SO(N)\times SO(N)$ when $N>2$.

Note that in \cite{Group}, it was found with the brutal force method that the cosets $SU(N)/SO(N)$ and $SO(2N)/SO(N)\times SO(N)$ possess the spin-4 quantum conserved currents. They used similar letters as ours in the counting. The crucial difference is that their letters $j^M$ are defined in the whole algebra while ours $k^\alpha$ only have the coset components. By lifting the letters with a conjugation\footnote{Basically, $j^M\sim gk^\alpha g^{-1}$ with $g\in G$.} into the whole algebra they can construct the gauge invariant operators from the trace operators.
As we explained in section 3, this counting is incomplete.

\subsection{$\mathbb{CP}^N$ coupled with fermions}
In this section, we generalize the KMS index to include fermionic letters. 
We have seen that the $\mathbb{CP}^{N-1}$ models are not quantum integrable. However it has been known for a while that the quantum integrability of the $\mathbb{CP}^{N-1}$ models can be restored by adding massless Dirac fermions \cite{Fermi}. To illustrate our construction, we focus on this model but our method is generally applicable.  

Without imposing the charge conjugation at the beginning, the KMS index can be computed in the presence of the fermionic letters.
The fermions are chiral so the possible letters are 
\bea 
D_-^m\psi_-,\quad  D_-^m\psi_-^\star, \quad D_+^m\psi_+,\quad D_+^m\psi_+^\star,
\eea 
which give rise to the character
\bea 
\chi_F=2\frac{\sqrt{qx}}{1-qx}+2\frac{\sqrt{q/x}}{1-q/x}.
\eea 
This character is problematic because in the conformal block the conformal dimension takes half-integer value such that the inversion formula does not work anymore. To cure this we can consider the ``bosonization" of the model by gauging the symmetry $U(1)\times U(1)$. For this gauge group  we introduce two more auxiliary parameters and modify the fermionic character as
\bea 
\chi_F(q,x,z_i)=\frac{\sqrt{qx}}{1-qx}(z_1+1/z_1)+\frac{\sqrt{q/x}}{1-q/x}(z_2+1/z_2).
\eea 
Recall the bosonic character is
\bea 
\chi_B(q,x,y_i)=(\frac{xq}{1-xq}+\frac{q/x}{1-q/x})(\sum_k y_k+\sum_k y_k^{-1}).
\eea 
Combining these two letters we can define the total partition function as a product $Z=Z_B Z_F$
\bea 
Z_F=\exp(\sum_{m=1} (-1)^{m+1}\frac{1}{m}\chi_F(q^m,x^m,z_i^m)),\quad Z_B=\exp(\sum_{m=1} \frac{1}{m}\chi_B(q^m,x^m,y_i^m))
\eea 
If we integrate out the auxiliary parameters $z_i,y_i$ we end up with the generating function without half-integer conformal block contributions because integrating out the gauge symmetry $U(1)\times U(1)$ guarantees the fermionic letters to group in pairs. 

 As argued the generating function will not contain unwanted characters corresponding to half-integer conformal dimensions. Note that we can not use the exponential form of the partition function to do this integral directly because it is not well-defined due to the appearance of the square root in the exponent.  Instead we should understand it as an expansion form so we introduce another parameter with respect to which we can do the expansion
\bea 
Z_F&=&\exp(\sum_{m=1} (-1)^{m+1}\frac{1}{m}u^m\chi_F(q^m,x^m,z_i^m)),\\
 Z_B&=&\exp(\sum_{m=1} \frac{u^m}{m}\chi_B(q^m,x^m,y_i^m)).
\eea
If we want to compute the index up to $J=6$, the expansion up to the power $u^{8}$ is enough. The resulted KMS indexes are
\bea 
\mathcal{I}(2)=-2,\quad \mathcal{I}(3)=-6,\quad  \mathcal{I}(4)=-12,\quad \mathcal{I}(5)=-26,\quad  \mathcal{I}(6)=-48,\dots
\eea 
Thus the GW argument fails. Now let us impose the charge conjugation symmetry.  In other words,  we need consider the charge conjugation invariant letters. The bosonic part can be treated in the same way. Gauging the $Z_2$ charge conjugation symmetry means that we should consider the real fermionic letters 
\bea 
D_-^m \psi_- D_-^m \psi_- ^\star,\quad D_+^m \psi_+ D_+^m \psi_+ ^\star.
\eea 
Even though these letters are bosonic, we need to take into account of the Pauli's exclusive principle
\bea 
(D_-^m \psi_- D_-^m \psi_- ^\star)^2=0.
\eea 
Therefore the partition function can be computed as
\bea\label{fpar} 
Z=\prod_{m=0}^\infty (1+(q x)^{2m+1}) (1+(q /x)^{2m+1})=[-q x,q^2x^2]_{\infty}[-q/ x,q^2/x^2]_{\infty},
\eea 
where the infinite products can be expressed with the $q$-pochhammer symbols. Using this fermionic partition function, one can find that  all the indexes are zero. The vanishing of the KMS index is due to the chiral structure.  Therefore, the final indexes \eqref{Discrete} are simply given by
\bea\label{CPF}
\mathcal{I}(2)=-1,\quad \mathcal{I}(3)=-3,\quad  \mathcal{I}(4)=-6,\quad \mathcal{I}(5)=-13,\quad  \mathcal{I}(6)=-24,\dots
\eea 
The negative KMS indexes show that the GW argument fails again.
\bigskip

In the literature, $\mathbb{CP}^{N-1}$ models are not often expressed as a coset model. Instead they are expressed in terms of complex vectors. We can also compute the KMS index in this formalism. The action is a complex version of \eqref{Naction}:
\bea \label{Caction}
 \mathcal{L}=\frac{1}{2\alpha} D_\mu n_i^\star D_{\mu} n^i,\quad \vec{n}\cdot\vec{n}^\star=1.
\eea 
The single-letters are
\bea 
&& P^{pq}_{++}=D_+^p \vec{n}^\star \cdot D_+^q \vec{n},\quad P^{pq}_{--}=D_-^p \vec{n}^\star \cdot D_-^q \vec{n},\nn
&&P^{pq}_{+-}=D_+^p \vec{n}^\star \cdot D_-^q \vec{n},\quad P^{pq}_{-+}=D_-^p \vec{n}^\star \cdot D_+^q \vec{n}
\eea 
and the corresponding character is given by
\bea 
\chi_B=2\frac{qx}{1-qx}\frac{qx^{-1}}{1-qx^{-1}}+(\frac{qx}{1-qx})^2+(\frac{q/x}{1-q/x})^2.
\eea 
If we want to impose the charge conjugation symmetry, the real single-letters are
\bea
P_{++}^{mm},\quad P_{--}^{mm},\quad P_{+-}^{mn}P_{-+}^{nm},\quad P_{++}^{mn}P_{++}^{nm},\quad P_{--}^{mn}P_{--}^{nm}
\eea
with the character
\bea 
\chi_B&=&G_s(q^2x^2)+G_s(q^2/x^2)+G_s(q^2x^2)G_s(q^2/x^2)\nn
&+&G_s(q^4 x^4)G_s(q^2 x^2)+G_s(q^4 /x^4)G_s(q^2 /x^2),\quad G_s(z)\equiv\frac{z}{1-z}.
\eea 
Combining with the ferminonic parts \eqref{fpar} we reproduce the exactly the same KMS indexes \eqref{CPF} for small $J<7$.

\section{Summary}
In this note, we elaborated the Komatsu, Mahajan and Shao's index of quantum integrability which systematized the analysis of Goldschmidt and Witten's argument. As applications, we revisited some quantum integrable coset models $\frac{SO(N)}{SO(N-1)}$,$\frac{SU(N)}{SO(N)}$ and $\frac{SO(2N)}{SO(N)\times SO(N)}$, and found the following results:
\begin{enumerate}
	\item The algebraic structure of the letters is crucial, particularly when it is trivial the KMS index vanishes for coset models.
	\item The KMS indexes of the $O(N)$ model in the coset description depend on $N$ when $N<7$. When $N\geq 7$ the KMS indexes will  be stable. After imposing the discrete symmetry, the KMS indexes become independent of $N$ and  predict the existences of spin-4 and spin-6 conserved currents.
	\item After imposing the discrete symmetries, the coset model $\frac{SO(4)}{SO(2)\times SO(2)}$ has KMS index $\mathcal{I}(4)=1$ suggesting the existence of a spin-4 conserved currents.
	\item The indexes of the coset models $\frac{SO(2N)}{SO(N)\times SO(N)}$ when $N\geq 3$ and  $\frac{SU(N)}{SO(N)}$  are all non-positive. The results are in conflict with the ones \cite{Group}. The reason is that in \cite{Group} the letters used in the counting have different algebraic structures from the ones of our letters.
\end{enumerate}

We also extended the KMS analysis to the theories with fermions and studied the $\mathbb{CP}^{N-1}$ model coupled with massless Dirac fermion. We found that KMS index in this kind of model failed to predict any high-spin conserved currents. Our analysis suggests that in order to have positive KMS index one has to  consider  coupling the fermions with non-trivial algebraic structure. For example, it would be interesting to consider the KMS index in the supersymmetric theories \cite{SUSY}.

\section*{Acknowledgments}
JT would like to thank Shota Komatsu for his inspiring lectures on integrability at the 13th Kavli Asian Winter School. The work was in part supported by NSFC Grant  No.~11335012, No.~11325522 and No. 11735001.

\end{document}